\documentclass[aps,twocolumn]{revtex4}
\usepackage [cp1251]{inputenc}
\usepackage[english]{babel}
\makeatletter\AtBeginDocument{\let\@elt\relax}\makeatother
\usepackage{amsmath}
\usepackage{bm}
\usepackage{graphicx}
\usepackage{color}
\usepackage[unicode=true,colorlinks=true,citecolor=blue,urlcolor=blue]{hyperref}

\newcommand{\braket}[1]{\left\langle #1 \right\rangle}

\renewcommand{\i}{{\rm i}}

\newcommand{\be}{\begin{equation}}
\newcommand{\ee}{\end{equation}}
\newcommand{\bea}{\begin{eqnarray}}
\newcommand{\eea}{\end{eqnarray}}

\newcommand {\e}{{\rm e}}
\renewcommand {\i}{{\rm i}}
\renewcommand {\d}{{\rm d}}

\graphicspath{{pics/}}

\begin{document}

\title{Theory of polarized photoluminescence of indirect band gap excitons\\ in type-I quantum dots}
\author{D.S.~Smirnov\footnote{smirnov@mail.ioffe.ru}, E.L. Ivchenko}
\affiliation{Ioffe Institute, 194021 St. Petersburg, Russia}

\begin{abstract}
  In this work, we theoretically investigate the optical orientation and alignment of excitons in quantum dots with weak electron-hole exchange interaction and long exciton radiative lifetimes. This particular regime is realized in semiconductor heterosystems where excitons are indirect in the ${\bm r}$ or ${\bm k}$ space. The main role in the fine structure of excitonic levels in these systems is played by the hyperfine interaction of the electron in the confined exciton and fluctuations of the Overhauser field. Along with it, the effects of nonradiative recombination and exchange interaction are considered. We start with the model of vanishing exchange interaction and nonradiative exciton recombination and then include them into consideration in addition to the strong Overhauser field. In the nanoobjects under study,  the polarization properties of the resonant photoluminescence are shown to vary with the external magnetic filed in completely different way as compared with the behaviour of the conventional quantum dot structures.
\end{abstract}
  \maketitle

  
\section{Introduction}

Excitons play a dominant role in optical properties of undoped and weakly doped low dimensional structures. The fine structure of excitonic levels is crucial for understanding and manipulating these properties. The fine structure of the exciton ground state in semiconductor quantum dots is determined by the splitting of four sublevels formed by the twofold degenerate electron and hole states~\cite{Ivchenko1995,Ivchenko1997a}. The nature of the splitting and the oscillator strength of optical transitions depend on the type of nanostructure. It is convenient to denote the four possible types of quantum dots as $d$-$r/d$-$k$, $ind$-$r/d$-$k$, $d$-$r/ind$-$k$ and $ind$-$r/ind$-$k$, where $d$ and $ind$ stand for {\it direct} and {\it indirect}, and $r$ and $k$ denote the real and reciprocal spaces. The quantum dots of the $d$-$r/d$-$k$ type were the first to be fabricated. The examples are CuCl and CdSe nanocrystals in glass matrices or colloidal solutions, and InGaAs/GaAs quantum dots fabricated by Stranski--Krastanov growth. The polarized exciton luminescence of this type of quantum dots is well studied~\cite{Ivchenko1998,Paillard2001,Koudinov2005,OOinGaN,InGaA2009}.

The $ind$-$r/d$-$k$ quantum dots are composed of two direct band gap semiconductors with the bottom of the conduction band and top of the valence band being located in different materials. As a result, the electron and hole wave functions are concentrated in different regions of the ${\bm r}$-space, but in the same region of the ${\bm k}$-space. Examples of such nanostructures are GaSb/GaAs quantum dots~\cite{GaSbII} and CdTe/CdSe, CdSe/ZnTe ``core-shell'' quantum dots~\cite{Kim,ASC2019} as well as transition metal dichalcogenides heterobilayers \cite{MoS2-WSe2, MoSe2-WSe2,Moire,MoireSmirnov}. The InAlAs/AlGaAs and Ga(As,P)/GaP heterosystems represent $ind$-$r/ind$-$k$-type quantum dots~\cite{GaAsP,InAlAs-AlGaAs}, also their role can be played by localized excitons in the GaAs/AlAs superlattice with thin GaAs layers. In such a multilayer structure, an electron in the exciton is localized within the AlAs layer in the $X$ valley of the Brillouin zone, while the hole is localized in the GaAs layer at the center of the Brillouin zone, this is the so-called $\Gamma$-$X$ exciton \cite{Guillaume,Ivchenko1991,Ivchenko1997,Shamirzaev}.

The (In,Al)As/AlAs heteropair represents quantum dots of the fourth type, namely $d$-$r/ind$-$k$~\cite{Nenashev,Shamirzaev2019,Smirnov2023}. Due to high AlAs potential barrier an electron and a hole are confined within the nanovolume of the (In,Al)As solid solution and also form a $\Gamma$-$X$ exciton. The less studied examples of $d$-$r/ind$-$k$ quantum dots are provided by GaAs/GaP, InSb/AlAs and GaSb/AlGaSb heteropairs~\cite{GaP,InSb,GaSb-AlGaSb}, and also effectively by width fluctuations of atomically thin (In,Al)(Sb,As)/AlAs and (Ga,Al)(Sb,As)/AlAs quantum wells~\cite{AlSb,GaSb}.

Under resonant optical excitation of quantum dots, the exciton spin dynamics is determined by the relation between the following parameters: the exchange interaction energy $\delta$, the radiative and nonradiative decay rates $\hbar/\tau_r, \hbar/\tau_{nr}$, the interaction energy $\varepsilon_N$ of the electron with a fluctuation of the nuclear field ${\bm B}_N$, and energy $\varepsilon_B$ of the Zeeman interaction with an external magnetic field ${\bm B}$. In ordinary $d$-$r/d$-$k$ quantum dots, the exchange interaction exceeds the other parameters and the influence of nuclei is insignificant for neutral excitons. By contrast, for charged quantum dots, the hyperfine interaction is the source of the spin relaxation both in the ground~\cite{Merkulov,Loss,Semenov} and  trion states~\cite{inertia_teor,inertia_exp}. In the GaAs/AlAs superlattice, a localized exciton encompasses a large number of nuclear spins $N$, and hence the hyperfine interaction $\varepsilon_N \propto N^{-1/2}$ is negligible.

In this paper, we theoretically study the optical orientation and alignment of excitons in quantum dots that are indirect in the $\bm r$ or $\bm k$ space. In this case, the reduced overlap of the electron and hole wave functions in the corresponding space suppresses the radiative recombination and the (long-range) exchange interaction making valid the following condition
\begin{equation} \label{cond}
\varepsilon_N \gg \delta, \frac{\hbar}{\tau_r}, \frac{\hbar}{\tau_{nr}}\:.
\end{equation}
To be specific we focus on (In,Al)As/AlAs quantum dots, which are the most studied, the generalization for the other systems is straightforward. Our aim is to examine the role of three small parameters on the right-hand side of the inequalities~\eqref{cond} on polarized exciton photoluminescence (PL) in an external magnetic field ${\bm B}$.

The article is structured as follows. In Sec.~\ref{II}, the states of the exciton quadruplet are introduced, their symmetry is analyzed for the point group D$_{2d}$, and the relationship between the polarizations of the secondary and initial radiation is found in case of strong splitting between sublevels of the quadruplet. In Sec.~\ref{IIa}, the simplest model is developed that neglects both the exciton nonradiative recombination and the exchange interaction. In this model, we calculate magnetic field dependencies of the degrees of circular and linear polarization of luminescence under polarized photoexcitation. The obtained results serve as a basis for the analysis of the roles of the nonradiative recombination in Sec.~\ref{III}, and the exchange interaction in Sec.~\ref{IV}. Section~\ref{V} summarizes and discusses the obtained results.

\section{Exciton quadruplet} \label{II}
To calculate the PL polarization dependencies on the parameters of the system, it is necessary to adopt a certain point symmetry of the quantum dot. For definiteness, we choose the symmetry of the point group D$_{2d}$. In this case, the electronic Bloch states in the conduction band,
\begin{equation} \label{electron}
\psi^{(e)}_{\frac12} = \alpha S, \psi^{(e)}_{-\frac12} = \beta S\:,
\end{equation} and hole states in the valence band, 
\begin{equation} \label{holev}
\psi^{(h)}_{\frac12} = - \beta \frac{X - {\rm i} Y}{\sqrt{2}}\:,\:\psi^{(h)}_{-\frac12} = \alpha \frac{X + {\rm i} Y}{\sqrt{2}}\:,
\end{equation}
transform under the operations $g$ of the group D$_{2d}$ according to the equivalent spinor representations $\Gamma_6$. Here $X,Y$ are the orbital Bloch functions at the center of the Brillouin zone which transform in D$_{2d}$ as the coordinates $x \parallel [100]$ and $y \parallel [010]$. 
  The electron orbital wave function $S$ mainly belongs to one of the $X$-valley states, while the optical properties are determined by its weak and spin independent mixing with the $s$-like ($\Gamma_1$) orbital of the conduction band at the center of the Brillouin zone~\cite{Willander}.
For the ground level of the exciton quadruplet in the quantum dot, we choose the basis of four states $\Psi_n$ ($n =1$--$4$) in the form of following products
\begin{equation} \label{psieh}
\psi^{(e)}_{\frac12} \psi^{(h)}_{-\frac12},~ \psi^{(e)}_{\frac12} \psi^{(h)}_{\frac12},~ \psi^{(e)}_{-\frac12} \psi^{(h)}_{-\frac12},~\psi^{(e)}_{-\frac12} \psi^{(h)}_{+\frac12}\:.
\end{equation} 
The states 2 and 3 are bright, the matrix elements for their optical excitation can be presented as
\begin{eqnarray} \label{Mpm}
&&  M^{({\rm abs})}_2 ({\bm e}^{(0)})= M_0 ( e^{(0)}_x - {\rm i} e^{(0)}_y)\:, \\
&&  M^{({\rm abs})}_3 ({\bm e}^{(0)})= M_0 ( e^{(0)}_x + {\rm i} e^{(0)}_y)\:, \nonumber
\end{eqnarray}
where ${\bm e}^{(0)}$ is the unit polarization vector of the exciting light. The coefficient $M_0$ describes the zero-phonon excitation of an electron-hole $X$-$\Gamma$ pair; it can be considered real-valued. For the nonresonant excitation the description remains valid provided the spin relaxation during the energy relaxation is negligible. The matrix elements for emission of a photon of the polarization ${\bm e}$ are related to $M^{({\rm abs})}$ by the relation
\[
M^{({\rm em})}_n ({\bm e}) \propto M^{({\rm abs})*}_n ({\bm e})~~(n = 2, 3)\:.
\]
The states 1 and 4 are optically dipole-inactive.

It is useful, instead of the basis states 2 and 3, to consider their linear combinations
\begin{equation} \label{xy}
\Psi_x = \frac{1}{\sqrt{2}} \left( \Psi_2 + \Psi_3 \right)\:,\:\Psi_y = \frac{{\rm i}}{\sqrt{2}} \left(\Psi_2 - \Psi_3\right)\:,
\end{equation}
which are optically active in the polarizations ${\bm e}^{(0)}, {\bm e} \parallel x$ and ${\bm e}^{(0)}, {\bm e} \parallel y$, respectively.

Below we consider the geometry of experiment where the exciting light and secondary radiation propagate along the structure growth axis $z$, and the light polarization vectors are lateral,
\begin{equation} \label{perp}
{\bm e}^{(0)} = (e^{(0)}_x, e^{(0)}_y,0), {\bm e} = (e_x, e_y, 0)\:.
\end{equation}
Then the components of polarization density matrices of the incident and emitted light, respectively $d^{(0)}_{\alpha, \beta}$ and $ d_{\alpha, \beta}$, are nonzero for $\alpha, \beta = x,y$ only.

Under the combined action of nuclear field fluctuations, the electron-hole exchange interaction and an external magnetic field, the exciton level splits into four sublevels $\varepsilon_j$ ($j= 1$--$4$). Let us decompose the exciton eigenstates $\Psi^{(j)}$ into the basis states $\Psi_1, \Psi_x, \Psi_y,\Psi_4$:
\begin{equation} \label{j-m}
\Psi^{(j)}= \sum\limits_{m = 1,x,y,4} C_m^{(j)} \Psi_m\:. 
\end{equation} 
The coefficients $C_m^{(j)}$ form a unitary matrix and satisfy the identities
\begin{equation} \label{CC}
\sum\limits_m C_m^{(j)} C_m^{(j')*} = \delta_{jj'}\:,\: \sum\limits_j C_m^{(j)} C_{m'}^{(j)*} = \delta_{mm'}\:. 
\end{equation}

We begin the description of polarized PL by considering the limit where the energy splittings between levels exceeds by far their natural width,
\begin{equation} \label{ineq}
|E_j - E_{j'}| \gg \frac{\hbar}{\tau_j}, \frac{\hbar}{\tau_{j'}}~~~(j' \neq j)\:,
\end{equation}
with $\tau_j$ being the exciton lifetime in the state $j$. Then the interference of quantum states is negligible and one can use the level populations $f_j$ instead of the exciton spin-density matrix. In this case, the polarization matrix of the emitted light has the form
\begin{equation} \label{dCC}
  d_{\alpha \beta} = \sum\limits_j d_{\alpha \beta}^{(j)} \propto M_0^2 \sum\limits_j C^{(j)*}_{\alpha}C^{(j)}_{\beta} f_j \:,
\end{equation} 
while the values of $f_j$ are found from
\begin{equation} \label{fj}
f_j \propto \tau_j \vert C^{(j)}_x e^{(0)}_x + C^{(j)}_y e^{(0)}_y \vert^2 \:.
\end{equation} 
The polarization-independent prefactors are omitted in Eqs.~\eqref{dCC} and~\eqref{fj}.
The inverse lifetimes of the excitonic sublevels contain two terms
\begin{equation} \label{tauf}
\frac{1}{\tau_j} = \frac{1}{\tau_r} \vert C^{(j)}_x + C^{(j)}_y \vert^2 + \frac{1}{\tau_{nr}}\:.
\end{equation}
They are determined by the radiative lifetime $\tau_r$ of the states \eqref{xy} and the nonradiative lifetime $\tau_{nr}$ which is independent of the index $j$.

The normalized Stokes parameters of the incident light are related with the polarization tensor $d^{(0)}_{\alpha \beta}$ by
\begin{eqnarray} \label{PPP}
&&P^{(0)}_l = \frac{d^{(0)}_{xx} - d^{(0)}_{yy}}{d^{(0)}_{xx} + d^{(0)}_{yy}}\:,\: P^{(0)}_{l'} = \frac{ 2 {\rm Re} \{d^{(0)}_{xy}\}}{d^{(0)}_{xx} + d^{(0)}_{yy}}\:, \\ && \hspace{1.5 cm} P^{(0)}_c = - \frac{2 {\rm Im} \{d^{(0)}_{xy}\}}{d^{(0)}_{xx} + d^{(0)}_{yy}}\:. \nonumber
\end{eqnarray}
We remind that the components $d^{(0)}_{\alpha \beta}$ are proportional to the products $e_\alpha^{(0)}e_\beta^{(0)*}$. By analogy with Eq.~(\ref{PPP}), it is convenient to introduce the combinations 
\begin{eqnarray}
&& \hspace{-4 mm} p^{(j)}_l = \frac{ \vert C_x^{(j)} \vert^2 - \vert C_y^{(j)} \vert^2}{\vert C_x^{(j) }\vert^2 + \vert C_y^{(j)}\vert^2}, p^{(j)}_{l'} = \frac{2{\rm Re} \{ C^{(j)}_x C^{(j)*}_y \} } {\vert C_x^{(j)} \vert^2 + \vert C_y^{(j)} \vert^2}, \nonumber\\ && \hspace{1.5 cm} p^{(j)}_c = - \frac{2{\rm Im} \{ C^{(j)}_x C^{(j)*}_y \}}{\vert C_x^{( j)}\vert^2 + \vert C_y^{(j)} \vert^2 }\:.
\end{eqnarray}
Then Eq.~\eqref{fj} takes the form
\begin{equation} \label{fj2}
f_j \propto \tau_j \Big( 1 + \sum\limits_{k= l, l', c} p^{(j)}_k P^{(0)}_k \Big)\:.
\end{equation}

Generally, the relation between the Stokes parameters of the secondary radiation $P_k$ and the parameters $P^{(0)}_{k' }$ can be written as
\begin{equation} \label{pk}
P_k = \sum_{k' = l, l', c} \Lambda_{kk'} P^{(0)}_{k'}\:.
\end{equation}
Neglecting nonradiative recombination processes ($\tau_{nr}/\tau_r \to \infty$) the PL intensity is independent of the polarization of the exciting light, and the matrix $\hat{\Lambda}$ is given by
\begin{equation} \label{Lambda}
\Lambda_{kk'} = \frac12 \sum\limits_{j=1}^4 ( \vert C^{(j)}_x \vert^2  + \vert C^{(j)}_y \vert^2)
p^{(j)}_k p^{(j)}_{k'}\:.
\end{equation} 

For excitons with large exchange splitting between $\Psi_{1,4}$ and $\Psi_{x,y}$ pairs of states [see Eqs.~\eqref{psieh} and~\eqref{xy}], the mixing between these pairs can be neglected, and it suffices for the calculation of the PL polarization to consider only the radiative doublet~\eqref{xy}~\cite{Ivchenko1998,bookIvchenko}. In this limit, the  summation in Eq.~(\ref{Lambda}) should be performed over the two radiative states $j=1,2$. Moreover, the sum $\vert C^{(j)}_x \vert^2 + \vert C^{(j)}_y \vert^2$ equals one and $p^{(1)}_k p^{(1)}_{k'} = p^{(2)}_k p^{(2)}_{k'}$.

\section{Neglecting the exchange interaction and nonradiative recombination} \label{IIa}
In this section we consider only the Zeeman interaction of electron and hole with external magnetic field and the hyperfine interaction of an electron with the host lattice nuclei. Then the exciton Hamiltonian takes the form
\begin{equation}   \label{eq:Ham_B}
  \mathcal H=\frac{\hbar}{2} \left[(\bm\Omega_L^e+\bm\Omega_N)\cdot {\bm \sigma}_e+\bm\Omega_L^h \cdot {\bm \sigma}_h \right]\:.
\end{equation}
Here ${\bm \sigma}_{e/h}$ are the electron/hole vectors of Pauli matrices,
$$\bm\Omega_L^e= \frac{\mu_B}{\hbar} \hat{g}_e \bm B\:,\quad \bm\Omega_L^h= \frac{\mu_B}{\hbar} \hat{g}_h \bm B$$
are the electron and hole spin precession frequencies in the magnetic field $\bm B$ with $\hat{g}_{e/h}$ being the tensors of the $g$ factors, $\mu_B$ the Bohr magneton and $\bm \Omega_N$ the electron spin precession frequency in the Overhauser field. The hole hyperfine interaction is suppressed because the hole Bloch functions $X,Y,Z$ vanish at the nuclear sites. In this work, we neglect phonon-assisted spin relaxation. The $D_{2d}$ symmetry allows for the anisotropy of the hole $g$ factor, which in the chosen basis has a simple form: $g_{h,zz} \equiv g_h^\parallel$, $g_{h,xx}=g_{h,yy} \equiv g_h^\perp$, and the other components are zero. Because of the weak spin-orbit coupling in X valleys, we neglect the anisotropy of the electron $g$ factor and set $g_{e,\alpha\beta}=g_e \delta_{\alpha\beta}$ with $\alpha,\beta=x,y,z$, $\delta_{\alpha\beta}$ being the Kronecker delta and $g_e\approx 2$.

The nuclear spin dynamics typically takes place at the time scale of $0.1$~ms~\cite{BookGlazov}. This may be comparable to the lifetime of dark excitons in (In,Al)As/AlAs quantum dots~\cite{Shamirzaev2003}, which may lead to peculiar features in the optical properties. However, we neglect the nuclear spin dynamics and assume $\bm\Omega_N$ to be ``frozen''~\cite{Merkulov}. The electron intravalley hyperfine interaction can be anisotropic~\cite{AlAs_hf}, but we neglect the anisotropy for simplicity as well as the intervalley hyperfine interaction~\cite{MX2_hf}. As a result, the distribution function of the electron spin precession frequencies takes the form
\begin{equation}
  \label{eq:F}
  \mathcal F(\bm\Omega_N)=\left(\sqrt{\frac{2}{\pi}}T_2^*\right)^3\e^{-2\left(\Omega_NT_2^*\right)^2},
\end{equation}
where the parameter $T_2^*$ characterizes the dispersion
\begin{equation} 
\langle \Omega_N^2 \rangle = \int \mathcal F(\bm\Omega_N) {\bm \Omega}_N^2 {\rm d} {\bm \Omega}_N = \frac 34 \frac{1}{T_2^{* 2}}\:,
\end{equation}
with ${\rm d} {\bm \Omega}_N={\rm d} \Omega_{N,x} {\rm d} \Omega_{N,y}{\rm d} \Omega_{N,z}$. Additionally, the parameter $T_2^*$ is a measure of the electron spin dephasing time. The product $\Omega_N T^*_2$ can take arbitrary values but typically it is of the order of unity.

According to the conditions~\eqref{cond}, the electron spin precession with the frequency $\bm\Omega_N$ is faster than the exciton recombination rate~\cite{DEP,DEP_details}. For this reason, the direction of $\bm \Omega_e = \bm\Omega_{L}^e+\bm\Omega_N$ fixes the quantization axis for the electron spin, and the electron eigen states are
\begin{eqnarray} \label{newstates}
&&\varphi^{(e)}_{\frac12} = {\rm e}^{-{\rm i} \varphi/2} \cos{\frac{\theta}{2}} \psi^{(e)}_{\frac12} + {\rm e}^{{\rm i} \varphi/2}\sin{ \frac{\theta}{2} } \psi^{(e)}_{- \frac12}\:,\\&&
 \varphi^{(e)}_{-\frac12} = - {\rm e}^{-{\rm i} \varphi/2} \sin{\frac{\theta}{2}} \psi^{(e)}_{\frac12} + {\rm e}^{{\rm i} \varphi/2}\cos{ \frac{\theta}{2} } \psi^{(e)}_{- \frac12}\:, \nonumber
\end{eqnarray}
where $\theta$ and $\varphi$ are the polar and azimuthal  angles of $\bm \Omega_e$. It is convenient to use the exciton spin-denisy matrix $\hat{\rho}$ in the basis $\Phi_{jm}=\varphi^{(e)}_j \psi^{(h)}_m$~($j,m=\pm1/2$), instead of the basis (\ref{psieh}). For zero exchange interaction,
the matrix $\hat{\rho}$ is block-diagonal in the electron spin subspace~\cite{Smirnov2023}:
\begin{equation} \label{lr}
\rho_{j'm',jm} = \delta_{j'j} \rho^{(j)}_{m',m}\:.
\end{equation}
The hole 2$\times$2 spin-density matrix can be conveniently presented in the form
\begin{equation} 
 \hat{\rho}^{(j)} = \left[ \begin{array}{cc} \frac{N^{(j)}}{2} + J^{(j)}_z & J^{(j)}_x - {\rm i} J^{(j)}_y\\ J^{(j)}_x + {\rm i} J^{(j)}_y & \frac{N^{(j)}}{2} - J^{(j)}_z \end{array}\right]\:,
\end{equation} 
where $N^{(j)}$ is the number of excitons with the electron spin $j$ and ${\bm J}^{(j)}$ is the hole average spin for the corresponding value of $j$.

We consider resonant exciton excitation with the rate $G$ and take into account a single decay channel in the system, namely, the radiative exciton recombination with the time $\tau_r$ defined in Eq.~\eqref{tauf}. In contrast to the previous section, the product $\Omega_L^h \tau_r$ can be arbitrary. The kinetic equations for $N^{(1/2)}$ and ${\bm J}^{(1/2)}$ take the form
\begin{widetext}
\begin{subequations}
  \label{eq:kin}
  \begin{eqnarray}
&&    \frac{{\rm d} N^{(\frac12)}}{{\rm d} t}=G\frac{1-P_c^{(0)}\cos\theta}{2}-\frac{1}{\tau_r}\left(\frac{N^{(\frac12)}}{2}+J^{(\frac12)}_z\cos\theta\right),\\
    &&\frac{{\rm d} J^{(\frac12)}_z}{{\rm d} t}=\Omega_{L,x}^hJ^{(\frac12)}_y-\Omega_{L,y}^hJ_x^{(\frac12)}+\frac{G}{4}\left(-P_c^{(0)}+\cos\theta\right)-\frac{1}{\tau_r}\left(\frac{J^{(\frac12)}_z}{2}+\frac{N^{(\frac12)}}{4}\cos\theta\right),\\
    &&\frac{{\rm d} J^{(\frac12)}_x}{{\rm d} t}=\Omega_{L,y}^hJ_z^{(\frac12)}-\Omega_{L,z}^hJ^{(\frac12)}_y+\frac{G}{4}\sin\theta \left( P_l^{(0)}\cos\varphi - P_{l'}^{(0)}\sin\varphi\right)-\frac{J^{(\frac12)}_x}{2\tau_r},\\
    &&\frac{{\rm d} J^{(\frac12)}_y}{{\rm d} t}=\Omega_{L,z}^hJ^{(\frac12)}_x-\Omega_{L,x}^hJ^{(\frac12)}_z-\frac{G}{4}\sin\theta\left(P_{l}^{(0)}\sin\varphi + P_{l'}^{(0)}\cos\varphi \right)-\frac{J^{(\frac12)}_y}{2\tau_r}\:,
  \end{eqnarray}
\end{subequations}
with $P_k^{(0)}~(k = l,l',c)$ being the Stokes parameters of the incident light, Eq.~(\ref{PPP}). The generation terms proportional to $G$ are derived taking into account that the wave function excited by the coherent light has the form $\Psi^{(0)}=e_x^{(0)}\Psi_x+e_y^{(0)}\Psi_y$. Under the cw excitation, the steady state is established and the time derivatives vanish. 

The intensities of circularly and linearly polarized photoluminescence in the $z$ direction read
\begin{subequations}
  \label{eq:Is}
  \begin{eqnarray}
&&    I_+^{(\frac12)}=\frac{N^{(\frac12)}-2J^{(\frac12)}_z}{2 \tau_r} (1 - \cos{\theta})\:,
    \qquad
    I_-^{(\frac12)}=\frac{N^{(\frac12)}+2J^{(\frac12)}_z}{2 \tau_r} (1 +\cos{\theta})\:,\\
&&    I_x^{(\frac12)}=\frac{1}{\tau_r}\left[\frac{N^{(\frac12)}}{2}+J^{(\frac12)}_z\cos\theta+\sin\theta\left(J^{(\frac12)}_x\cos\varphi-J^{(\frac12)}_y\sin\varphi\right)\right]\:,
    \\
&&    I_y^{(\frac12)}=\frac{1}{\tau_r}\left[\frac{N^{(\frac12)}}{2}+J^{(\frac12)}_z\cos\theta-\sin\theta\left(J^{(\frac12)}_x\cos\varphi-J^{(\frac12)}_y\sin\varphi\right)\right]\:, \nonumber \\
&&    I_{x'}^{(\frac12)}=\frac{1}{\tau_r}\left[\frac{N^{(\frac12)}}{2}+J^{(\frac12)}_z\cos\theta-\sin\theta\left(J^{(\frac12)}_x\sin\varphi+J^{(\frac12)}_y\cos\varphi\right)\right]\:,
    \\
&&    I_{y'}^{(\frac12)}=\frac{1}{\tau_r}\left[\frac{N^{(\frac12)}}{2}+J^{(\frac12)}_z\cos\theta+\sin\theta\left(J^{(\frac12)}_x\sin\varphi+J^{(\frac12)}_y\cos\varphi\right)\right]\:. \nonumber
  \end{eqnarray}
\end{subequations}
\end{widetext}
Equations for $N^{(-1/2)}$, ${\bm J}^{(-1/2)}$ and the corresponding intensities have the same form as Eqs.~\eqref{eq:kin} and~\eqref{eq:Is}, but with the replacement of $\cos{\theta}$ and $\sin{\theta}$ with $-\cos{\theta}$ and $-\sin{\theta}$.

\subsection{Longitudinal magnetic field (Faraday geometry)}

In the longitudinal magnetic field, the substitution of the solution of Eqs.~\eqref{eq:kin} into Eqs.~\eqref{eq:Is} yields
\begin{equation} \label{Llong}
  \hat{\Lambda}=
    \begin{pmatrix}
    \frac{\sin^2\theta}{1+(2\Omega_{L,z}^h\tau_r)^2} & \frac{2\Omega_{L,z}^h\tau_r\sin^2\theta}{1+(2\Omega_{L,z}^h\tau_r)^2} & 0 \\
    -\frac{2\Omega_{L,z}^h\tau_r\sin^2\theta}{1+(2\Omega_{L,z}^h\tau_r)^2} & \frac{\sin^2\theta}{1+(2\Omega_{L,z}^h\tau_r)^2} & 0 \\
    0 & 0 & 1
  \end{pmatrix}.
\end{equation}
The multiplier $2$ before $\Omega_{L,z}^h\tau_r$ reflects the fact that for the fast hole spin precession the bright excitons are equally mixed with the dark ones. This expression should be averaged over the random Overhauser fields as follows
\begin{equation}
\langle \hat{\Lambda} \rangle=\int {\hat \Lambda} {\cal F}({\bm \Omega}_N){\rm d} {\bm \Omega}_N\:.
\end{equation}
The integration can be readily done using the equation~\cite{Petrov2008,PRC}
\begin{multline} \label{Dawson}
  \braket{\sin^2\theta}=\frac{1}{2\left(\Omega_L^{e}T_2^*\right)^2}\left[1-\frac{1}{\sqrt{2}\Omega_L^eT_2^*}D\left(\sqrt{2}\Omega_L^eT_2^*\right)\right]\\\approx\frac{2}{3}\frac{1}{1+\left(\Omega_L^{e}T_2^*\right)^2}\:, \hspace{2 cm}
\end{multline}
where $D(x)=\exp(-x^2)\int_0^x\exp(y^2) {\rm d} y$ is the Dawson integral. The approximation reproduces the exact function up to $4\%$.

\begin{figure}
  \centering
  \includegraphics[width=0.95\linewidth]{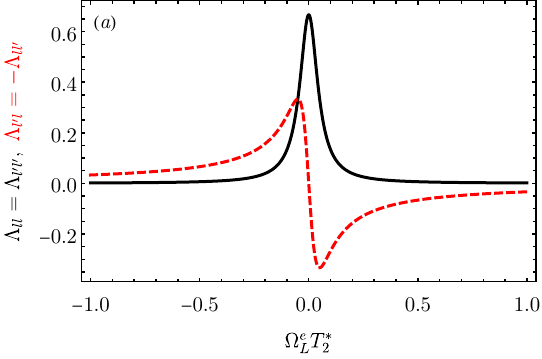}
  \includegraphics[width=0.95\linewidth]{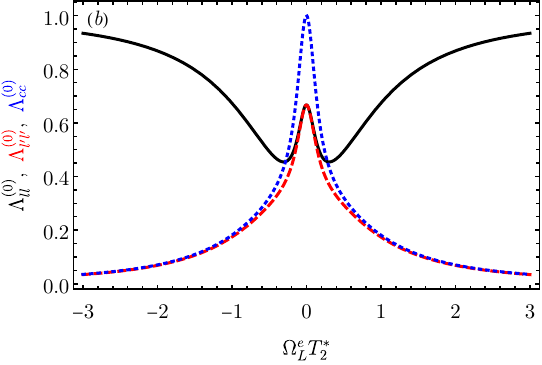}
  \caption{
    \label{fig:components}
        (a) Coefficients $\Lambda_{ll}=\Lambda_{l'l'}$ (black solid line) and $\Lambda_{l'l}=-\Lambda_{ll'}$ (red dashed line), calculated after Eq.~\eqref{eq:L_Farad} as the functions of the longitudinal magnetic field for $g_h^\parallel\tau_r=10g_eT_2^*$.
    (b) Diagonal components of the matrix $\hat{\Lambda}^{(0)}$ calculated after Eq.~\eqref{eq:La} as functions of the transverse magnetic field for $g_h^\perp\tau_r=3g_eT_2^*$. The exciton optical alignment along and across the magnetic field, $\Lambda_{ll}^{(0)}$, is shown by the black solid curve; the alignment in the axes rotated by $\pi/4$, $\Lambda_{l'l'}^{(0)}$, is shown by the red dashed curve, and the optical orientation, $\Lambda^{(0)}_{cc}$ is shown by the blue dotted curve.}
\end{figure}

In the realistic case of $\Omega_L^eT_2^*\ll\Omega_L^h\tau_r$ or $g_eT_2^*\ll g_h^\parallel\tau_r$ (we remind that $1/\Omega_N \sim T_2^*\ll\tau_r$), we obtain from Eqs.~(\ref{Llong})--(\ref{Dawson})
\begin{equation}
\langle \hat{\Lambda} \rangle=
  \begin{pmatrix}
    \frac{2/3}{1+(2\Omega_{L,z}^h\tau_r)^2} & \frac{(4/3)\Omega_{L,z}^h\tau_r}{1+(2\Omega_{L,z}^h\tau_r)^2} & 0 \\
    -\frac{(4/3)\Omega_{L,z}^h\tau_r}{1+(2\Omega_{L,z}^h\tau_r)^2} & \frac{2/3}{1+(2\Omega_{L,z}^h\tau_r)^2} & 0 \\
    0 & 0 & 1
  \end{pmatrix}.
  \label{eq:L_Farad}
\end{equation}
The two different nontrivial components of this matrix are shown in Fig.~\ref{fig:components}(a).
\vspace{2 mm}

\subsection{Transverse magnetic field (Voight geometry)}
In the transverse magnetic field, the nonzero components of the hole spin precession frequency ${\bm \Omega}^h_L$ are $\Omega_{L,x}^h=\Omega_L^h\cos\alpha$, and $\Omega_{L,y}^h=\Omega_L^h\sin\alpha$, where $\alpha$ is the angle between $\bm\Omega_L^h$ and the $x$ axis. Then we obtain from Eqs.~\eqref{eq:kin} and~\eqref{eq:Is}
\begin{widetext}
\begin{eqnarray}
  \label{eq:Lambda_init}
  {\hat\Lambda}=
  \begin{pmatrix}
    \sin^2\theta\cos^2(\varphi+\alpha) & -\frac{1}{2}\sin^2\theta\sin(2\varphi+2\alpha) & 0 \\
    -\frac{1}{2}\sin^2\theta\sin(2\varphi+2\alpha) & \sin^2\theta\sin^2(\varphi+\alpha) & 0 \\
    0 & 0 & \cos^2\theta
  \end{pmatrix}
  +\ \frac{\sin^4\theta}{\sin^2\theta+(2\Omega_{L}^h\tau_r)^2}
  \begin{pmatrix}
    \sin^2(\varphi+\alpha) & \frac{1}{2}\sin(2\varphi+2\alpha) & 0 \\
    \frac{1}{2}\sin(2\varphi+2\alpha) & \cos^2(\varphi+\alpha) & 0 \\
    0 & 0 & 1
  \end{pmatrix}\:. \nonumber
\end{eqnarray}
Averaging the matrix ${\hat \Lambda}$ in the same way as in the previous subsection we obtain 
\begin{eqnarray}
  \label{eq:Lxxx}
  && \langle \hat{\Lambda} \rangle=
  \frac{1}{(\Omega_L^eT_2^*)^2+1}
  \begin{pmatrix}
    \frac{1}{3}+\cos^2(2\alpha)(\Omega_L^eT_2^*)^2 & -\frac{1}{2}\sin(4\alpha)(\Omega_L^eT_2^*)^2 & 0 \\
    -\frac{1}{2}\sin(4\alpha)(\Omega_L^eT_2^*)^2 & \frac{1}{3}+\sin^2(2\alpha)(\Omega_L^eT_2^*)^2 & 0 \\
    0 & 0 & \frac{1}{3} \\
  \end{pmatrix}
       +\frac{1/3}{1+15(\Omega_{L}^h\tau_r)^2/3}
  \begin{pmatrix}
    1 & 0 & 0 \\
    0 & 1 & 0 \\
    0 & 0 & 2
\end{pmatrix}\:.  \nonumber  
\end{eqnarray}
\end{widetext}
Notably, this expression reveals an anisotropy of the QD having the D$_{2d}$ symmetry. This anisotropy originates from the fixed orbitals $X,Y$ in the basis~\eqref{holev} which defines the selection rules for linearly polarized light and the heavy-hole Zeeman Hamiltonian~\eqref{eq:Ham_B}~\cite{Pikuses}. As a result, the polarization of the emitted light depends not only on the angle between polarization of the exciting light and magnetic field, but also on their orientation with respect to the crystallographic axes.

In the limit of strong magnetic field, $\Omega_L^eT_2^*\gg 1$, we obtain
\begin{equation}
  \langle \hat{\Lambda} \rangle=
  \begin{pmatrix}
    \cos^2(2\alpha) & -\frac{1}{2}\sin(4\alpha) & 0 \\
    -\frac{1}{2}\sin(4\alpha) & \sin^2(2\alpha) & 0 \\
    0 & 0 & 0 \\
  \end{pmatrix}.
\end{equation}
First, we note that this expression can be obtained from Eq.~\eqref{eq:Lambda_init} using $\theta=\pi/2$ and $\phi=\alpha$. Second, it can be obtained from Eq.~\eqref{Lambda} from the previous section using the eigenfunctions
\begin{subequations} \label{strong}
  \begin{eqnarray} 
    && \Psi^{(1)}=\left(\Psi_2+\e^{\i\alpha}\Psi_1+\e^{\i\alpha}\Psi_4+\e^{2\i\alpha}\Psi_3\right)/2,\\
    && \Psi^{(2)}=\left(\Psi_2-\e^{\i\alpha} \Psi_1 +\e^{\i\alpha} \Psi_4-\e^{2\i\alpha}\Psi_3\right)/2,\\
    && \Psi^{(3)}=\left(\Psi_2 + \e^{\i\alpha}\Psi_1 - \e^{\i\alpha}\Psi_4 - \e^{2\i\alpha}\Psi_3\right)/2,\\
    && \Psi^{(4)}=\left(\Psi_2-\e^{\i\alpha}\Psi_1-\e^{\i\alpha}\Psi_4+\e^{2\i\alpha}\Psi_3\right)/2,
  \end{eqnarray}
\end{subequations}
which give the coefficients $|C_x^{(j)}|^2 + |C_y^{(j)}|^2 = 1/2$ for all $j$ and polarizations $p_{l}^{(1)} = p_l^{(4)} = - p_{l}^{(2)} = - p_l^{(3)} = \cos{(2\alpha)}$, $p_{l'}^{(2)} = p_{l'}^{(3)}= - p_{l'}^{(1)} = - p_{l'}^{(4)} = \sin{(2\alpha)}$, and $p_c^{(j)}=0$ for all $j$.

In the following, we focus on the specific case of $\alpha=0$. Denoting the matrix $\hat{\Lambda}$ at $\alpha = 0$ as $\hat{\Lambda}^{(0)}$ we have
\begin{multline}
\langle \hat{\Lambda}^{(0)} \rangle=
  \begin{pmatrix}
    \frac{(\Omega_L^eT_2^*)^2+1/3}{(\Omega_L^eT_2^*)^2+1} & 0 & 0 \\
    0 & \frac{1/3}{(\Omega_L^eT_2^*)^2+1} & 0 \\
    0 & 0 & \frac{1/3}{(\Omega_L^eT_2^*)^2+1} \\
  \end{pmatrix}
  \\
  +\frac{1/3}{1+15(\Omega_{L}^h\tau_r)^2/3}
  \begin{pmatrix}
    1 & 0 & 0 \\
    0 & 1 & 0 \\
    0 & 0 & 2
  \end{pmatrix}.
  \label{eq:La}
\end{multline}
In this case, there is no polarization conversion. The three diagonal components of the matrix $\hat{\Lambda}^{(0)}$ are shown in Fig.~\ref{fig:components}(b) as functions of the magnetic field. Notably for $\alpha$ being a multiple of $\pi/4$, the matrix $\braket{\hat\Lambda}$ has the same form in the rotated coordinate frame $x', y', z$ with the $x'$ axis parallel to the direction of the magnetic field.

\section{Role of nonradiative recombination} \label{III}

Let us take into account the nonradiative recombination time $\tau_{nr}$ satisfying the condition~\eqref{cond}. Then one should replace $\d/\d t$ with $\d/\d t+1/\tau_{nr}$ in the kinetics equations~\eqref{eq:kin}. At the same time, in the expressions for the PL intensities  (\ref{eq:Is}), the common factor $\tau_r^{-1}$ remains unchanged. As a consequence, the PL intensity becomes dependent on the magnetic field, and this is the most important manifestation of the nonradiative recombination channel.

\begin{figure}
  \centering
  \includegraphics[width=0.95\linewidth]{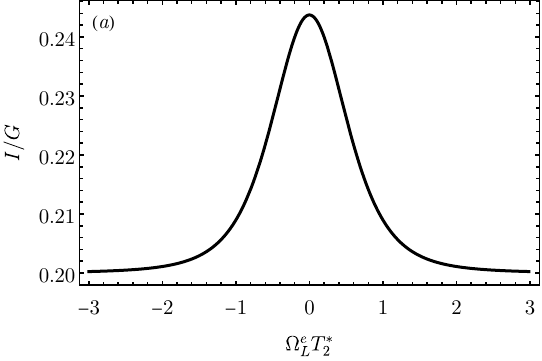}
  \includegraphics[width=0.95\linewidth]{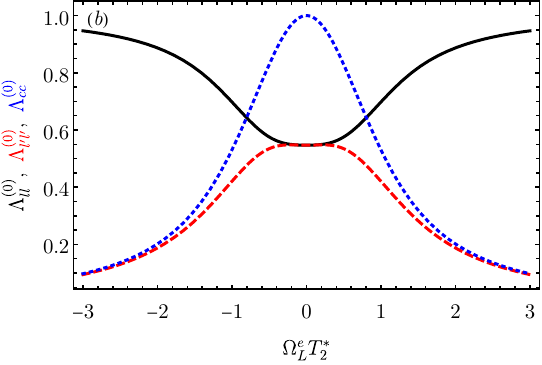}
  \includegraphics[width=0.95\linewidth]{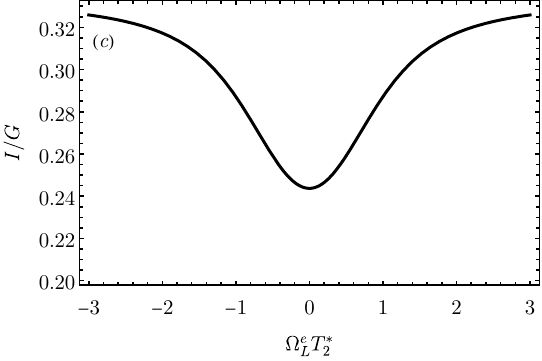}
  \caption{
    \label{fig:nonradiative}
The PL intensity (a) and polarization (b) as a function of the transverse magnetic field ${\bm B} \parallel x$. The solid, dashed, and dotted curves are the degrees of polarization $P_l, P_{l'}$ and $P_c$ under excitation by light of the corresponding polarization $P^{(0)}_l = 1, P^{(0)}_{l'} = 1$, and $P^{(0)}_c = 1$. (c) Magnetic field dependence of the PL intensity at ${\bm B} \parallel z$.}
\end{figure}

Fig.~\ref{fig:nonradiative} shows the PL intensity and polarization as functions of the transverse ($a,b$) and longitudinal ($c$) magnetic field calculated for $\tau_{nr}=\tau_r/2$ and the same other parameters as in Fig.~\ref {fig:components}. In the absence of magnetic field, the total PL intensity $$I=I_+^{(1/2)}+I_-^{(1/2)}+I_+^{(-1/2)}+I_-^{(-1/2)}$$ is given by
\begin{equation}
  I =G\left[1-2\frac{\tau_r}{\tau_{nr}}+2\frac{\tau_r^2}{\tau_{nr}^2}\ln\left(\frac{\tau_r+\tau_{nr}}{\tau_r}\right)\right]\:.
\end{equation}

In a longitudinal magnetic field $B_z$, due to the time-reversal symmetry, the intensity $I$ is an even function of $B_z$. Therefore, this function is an invariant of the symmetry group D$_{2d}$ (representation $\Gamma_1$) and does not depend on the excitation polarization since the differences $I_x - I_y$, $I_{x'} - I_{ y'}$, $I_+ - I_-$, $I_k$ being $I_k^{(1/2)}+I_k^{(-1/2)}$, belong to the representations $\Gamma_3, \Gamma_4, \Gamma_2 \neq \Gamma_1$ respectively. As a result, the curve in Fig.~\ref{fig:nonradiative}(c) does not depend on the excitation polarization. It shows an increase in the PL intensity up to
\begin{equation}
  \label{eq:Iz}
  I= \frac{G\tau_{nr}}{\tau_r + \tau_{nr}}
\end{equation}
with the increasing longitudinal magnetic field.

In a transverse field, the dependence $I(B_x)$ is also even in $B_x$. It is convenient to present the square $B_x^2$ as the invariant $(B_x^2 + B_y^2)/2$ and the difference $(B_x^2 - B_y^2)/2$ that transforms according to the representation $\Gamma_3$. It follows then that the intensity $I$ is insensitive to the Stoke parameters $P^{(0)}_{l'}$ and $P_c^{(0)}$, but can depend on $P_l$. The calculation, however, shows that under the condition $|g_h^\perp/g_e|\ll1$ the dependence of $I$ on $P^{(0)}_l$ vanishes, and the dependence of $I(B_x)$ is the same for all excitation polarizations, Fig.~\ref{fig:nonradiative}(a). In a strong transverse magnetic field, the intensity decreases to
\begin{equation}
I=\frac{G\tau_{nr}}{2\tau_r+\tau_{nr}}\:.
\end{equation}

The magnetic field dependencies of polarizations vary continuously with the decreasing time $\tau_{nr}$. The remarkable peak-to-valley ratio of the $P_l(B_z)$ curve in Fig.~\ref{fig:components}(b) decreases with increasing role of the nonradiative recombination, and for $\tau_{nr} = 0.5 \tau_r$ the peak disappears.

\section{Role of exchange interaction} \label{IV}

\begin{figure}
  \centering
  \includegraphics[width=0.95\linewidth]{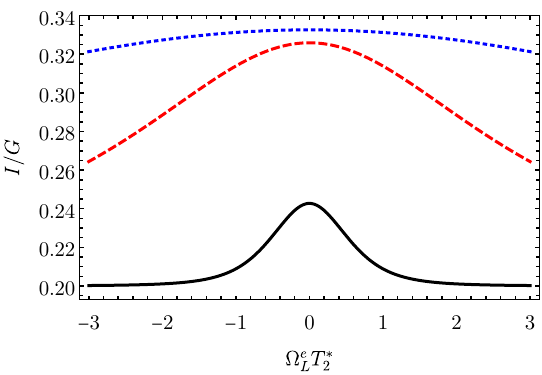}
  \caption{\label{fig:exchange_int}
    Dependence of the PL intensity on the magnetic field applied in the Voight geometry for the different exchange interaction strengths $\delta_0T_2^*=0$ (black solid curve), $3$ (red dashed curve), and $10$ (blue dotted curve).
  }
\end{figure}

The short-range exchange interaction between an electron and a hole is described by the Hamiltonian~\cite{Ivchenko1997a}
\begin{equation}
  \label{eq:exch}
  V_{\rm exch}= \hbar \delta_0 \sigma_{e,z}\sigma_{h,z}/2,
\end{equation}
where $\delta_0$ is the splitting between dark and bright excitonic states. Fig.~\ref{fig:exchange_int} shows the modification of the PL intensity dependence on the transverse magnetic field by the exchange interaction. It is calculated numerically using the spin density matrix formalism with the same parameters as for Fig.~\ref{fig:nonradiative}(a) and $T_2^*/\tau_r=0.005$. The total intensity weakly depends on the polarization of the excitation, so we show the results for the unpolarized excitation to be specific. One can see that increase of the exchange interaction leads to the disappearance of this dependence due to the weakened mixing between bright and dark excitons. This takes place at $\delta_0\sim 1/T_2^*$ and in the limit of $\delta_0T_2^*\gg 1$ the intensity saturates at the value given by Eq.~\eqref{eq:Iz} similarly to the strong longitudinal magnetic field.

\begin{figure}
  \centering
  \includegraphics[width=0.95\linewidth]{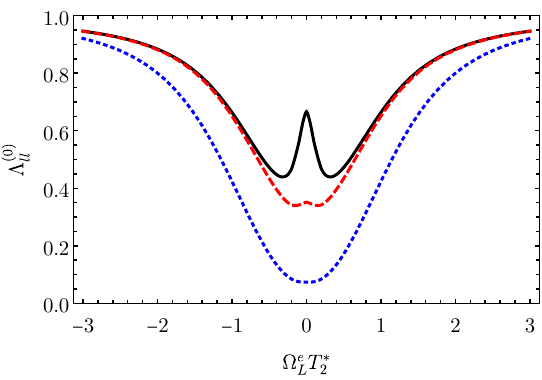}
  \caption{\label{fig:exchange_pol}
    Dependence of the PL linear polarization degree $\Lambda_{ll}^{(0)}$ on the magnetic field applied in the Voight geometry for the different exchange interaction strengths $\delta_0\tau_r=0$ (black solid curve), $1.5$ (red dashed curve), and $10$ (blue dotted curve). Note the difference in the scale of the exchange interaction with Fig.~\ref{fig:exchange_int}. The alignment monotonously decreases with increase in $\delta_0$.}
\end{figure}

However, a much weaker exchange interaction $\delta_0\sim1/\tau_r$ can strongly modify the polarization of the PL. This effect can be described by Eqs.~\eqref{eq:kin} and~\eqref{eq:Is} with the only replacement $\Omega_{L,z}^h \to \Omega_{L,z}^h + \delta_0 \cos{\theta}$, which can be seen from Eq.~\eqref{eq:exch}. The result of the calculation is shown in Fig.~\ref{fig:exchange_pol} for the same parameters as in Fig.~\ref{fig:components} (without nonradiative recombination) and different strengths of the exchange interaction. One can see that the increase of the exchange interaction reverses the narrow component of $\Lambda_{ll}^{(0)}(B_x)$: a dip in the polarization becomes deeper. This effect can be described analytically similarly to Sec.~\ref{IIa}, but the result turns out to be too cumbersome.

\section{Discussion and conclusion} \label{V}

Figures \ref{fig:components}--\ref{fig:exchange_pol} illustrate the characteristics of resonant polarized PL spectroscopy of $d$-$r/ind$-$k$ quantum dots. Due to small exchange splitting in these dots, the fluctuations of the Overhauser field have a significant effect on the electron spin. As a consequence, at zero magnetic field in the absence of exchange interaction and spin-lattice relaxation, the degree of PL linear polarization excited by linearly polarized light decreases to $2/3$, see Eqs.~\eqref{eq:L_Farad} and~\eqref{eq:Lxxx} for $\langle \hat{\Lambda} \rangle$. Inclusion of a weak short-range exchange interaction~\eqref{eq:exch} results in an additional reduction of the exciton alignment which has a double-scale character. At the scale $\delta_0 \sim 1/\tau_r \ll 1/T^*_N$, the electron spin states (\ref{newstates}) are unaffected by the exchange interaction but the electron occupying these states partially depolarizes the hole spin. As $\delta_0$ increases to 10$\tau_r^{-1}$, the linear polarization at zero magnetic field decreases from 2/3 to 0.1.

If $\delta_0$ prevails over the nuclear fluctuations $\sqrt{\langle \Omega^2_N \rangle} \propto 1/T_2^*$, the exchange interaction suppresses the influence of the lateral component ${\bm \Omega}_N$ of the nuclear field, thereby increasing the effect of the longitudinal component $\Omega_{N,z}$ on the electron. Note that, in $d$-$r/d$-$k$ quantum wells, the radiative damping $\tau_r^{-1}$ exceeds $\Omega_N$ by far, the nuclear field has no time to produce a depolarizing influence and the degree of linear polarization $P_l$ or $P_{l'}$ may reach 100\%.

Application of the longitudinal magnetic field results in a hole spin precession which, in its turn, leads to the conversion of linear polarizations $P_l \leftrightarrow P_{l'}$ and an overall decrease of optical alignment, dashed and solid curves in Fig.~\ref{fig:components}(a), similarly to the polarization behaviour in conventional $d$-$r/d$-$k$ quantum dots. 

An external transverse magnetic field leads to the mixing of hole spin-up and spin-down states controlled by the transverse $g$-factor $g_h^{\perp}$. This results in a partial suppression of the optical orientation. When the magnetic field reaches the value of typical Overhauser field fluctuations, the electron spin states also get affected. With further increase of the field, the optical orientation vanishes, as shown in Fig.~\ref{fig:components}(b). The field-induced mixing of the spin states also suppresses the optical alignment component $\Lambda_{l',l'}^{(0)}$.

The alignment component $\Lambda_{l,l}^{(0)}$ shows quite different behaviour, the black solid curve in Fig.~\ref{fig:components}(b). At $\Omega_L^e T^*_2 < 1$, the electron spin states (\ref{newstates}) are unaffected by the magnetic field while the hole spin states are depolarized at $\Omega_L^h \tau_r \ge 1$, and the polarization $P_l$ decreases with the increasing field. However, the strong magnetic field, $\Omega_L^e T^*_2 \gg 1$, suppresses the nuclear field and tends to form both the electron and hole eigenstates with the spins parallel or antiparallel to ${\bm B}$. As a result, in the geometry ${\bm B} \parallel x$ among four exciton split sublevels (\ref{strong}) two are active in the $x$ polarization, the remaining two are $y$-polarized, and one has $P_l = P^{(0)}_l$. 

The nonradiative exciton recombination manifests itself by expected dependence of the PL intensity on the transverse magnetic field, Fig.~\ref{fig:nonradiative}(a). The larger magnetic field, the stronger mixing between bright and dark states, and the smaller the intensity. The longitudinal magnetic field acts in the opposite way: it cancels out the effect of the nuclear field and decouples the bright and dark excitons causing growth of the PL intensity, as shown in Fig.~\ref{fig:nonradiative}(c). Qualitatively, the dependencies of $P_l, P_{l'}$ and $P_c$ on the transverse magnetic field remain the same as in Fig.~\ref{fig:components}(b) obtained in the absence of nonradiative recombination, but they get smoother. In particular, the peak with two symmetrical valleys in the component $\Lambda_{ll}^{(0)}$ disappears. It happens because the width of this peak is determined by the exciton lifetime, and the nonradiative recombination reduces this time thus broadening the peak, which leads to its disappearance for the chosen parameters.

The strong exchange splitting, $\delta_0 \ge 1/T_2^*$, suppresses mixing between the bright and dark exciton states. As a result, the effect of the transverse magnetic field on the PL intensity is reduced, and the dependence $I(B)$ becomes weaker and broader, as one can see in Fig.~\ref{fig:exchange_int}. 

Finally, it is noteworthy to mention that here we have neglected the magnetic dipole-dipole hole-nuclear interaction assuming the corresponding hole spin precession frequency $\Omega_N^h$ to be small compared to the exciton inverse lifetime. Because of the long lifetimes of excitons in $d$-$r/ind$-$k$ quantum dots, this assumption may be violated. An analysis of the model where $\Omega_N^h$ is nonzero will be reported elsewhere.

Also, it would be interesting to generalize the developed theory for quantum dots of the C$_{2v}$ symmetry, where the two different contributions to the transverse hole $g$ factor are symmetry allowed. The role of intervalley scattering in (In,Al)As/AlAs quantum dots is not clear yet. For moir\'e excitons in bilayers of transition metal dichalcogenides the ground electron and hole states form a quadruplet of excitons with the selection rules like those in Eq.~\eqref{Mpm}. With this in mind, we can predict the similar physics except for the much smaller transverse electron and hole $g$ factors, which describe intervalley mixing.

In conclusion, we have developed a theory of polarized photoluminescence of excitons confined in quantum dot structures with weak electron-hole exchange interaction. The particular nanosystem can be realized in quantum dots where the excitons are indirect either in the real or reciprocal space. For them the leading role is played by the electron-nuclear hyperfine interaction. We have derived a relation between the PL Stokes parameters and those of the exciting light. It is quite different from the similar relation for conventional nanoobjects with excitons characterized by a strong exchange interaction and short lifetimes. In particular, the optical alignment can be a nonmonotonic function of the transverse magnetic field. We have started with the model of vanishing exchange interaction and nonradiative exciton recombination and then included them into consideration. We believe that the present work opens up new possibilities in the exciton physics and can be applied to study the confined excitons in (In,Al)As/AlAs quantum dots, interlayer excitons in transition metal dichalcogenides etc.

\section*{Acknowledgments}
We thank Yu.G. Kusraev, S.V. Nekrasov and T.S. Shamirzaev for fruitful discussions and Foundation for the Advancement of Theoretical Physics and Mathematics ``BASIS''. The work is supported by the Russian Science Foundation Grant no. 23-12-00142.

\end{document}